\documentclass[preprint,journal]{vgtc}            

\newcommand{\cells}{\textsc{Cells}}
\newcommand{\cortex}{\textsc{Cortex}}
\newcommand{\fiber}{\textsc{Fiber}}

\onlineid{1392}

\ieeedoi{10.1109/TVCG.2023.3326573}

\vgtccategory{Data Transformations}

\vgtcpapertype{algorithm/technique}

\title{Fast Compressed Segmentation Volumes for Scientific Visualization}

\author{%
  \authororcid{Max Piochowiak}{0000-0003-1980-6146}, and 
  \authororcid{Carsten Dachsbacher}{0000-0003-4690-3574}
}

\authorfooter{
   \item
   	Max Piochowiak and Carsten Dachsbacher are with Karlsruhe Institute of Technology.
   	E-mail:  \{max.piochowiak\,$|$\,dachsbacher\}@kit.edu\,.
}

\abstract{
Voxel-based segmentation volumes often store a large number of labels and voxels, and the resulting amount of data can make storage, transfer, and interactive visualization difficult. We present a lossless compression technique which addresses these challenges. It processes individual small bricks of a segmentation volume and compactly encodes the labelled regions and their boundaries by an iterative refinement scheme. The result for each brick is a list of labels, and a sequence of operations to reconstruct the brick which is further compressed using rANS-entropy coding.
As the relative frequencies of operations are very similar across bricks, the entropy coding can use global frequency tables for an entire data set which enables efficient and effective parallel (de)compression.
Our technique achieves high throughput (up to gigabytes per second both for compression and decompression) and strong compression ratios of about 1\% to 3\% of the original data set size while being applicable to GPU-based rendering. 
We evaluate our method for various data sets from different fields and demonstrate GPU-based volume visualization with on-the-fly decompression, level-of-detail rendering (with optional on-demand streaming of detail coefficients to the GPU), and a caching strategy for decompressed bricks for further performance improvement.
}


\teaser{
  \centering
  \setlength{\unitlength}{\columnwidth}
  \relax%
  \begin{picture}(1,0.5695)%
    \setlength\tabcolsep{0pt}%
    \put(0,0){\includegraphics[width=\unitlength,page=1]{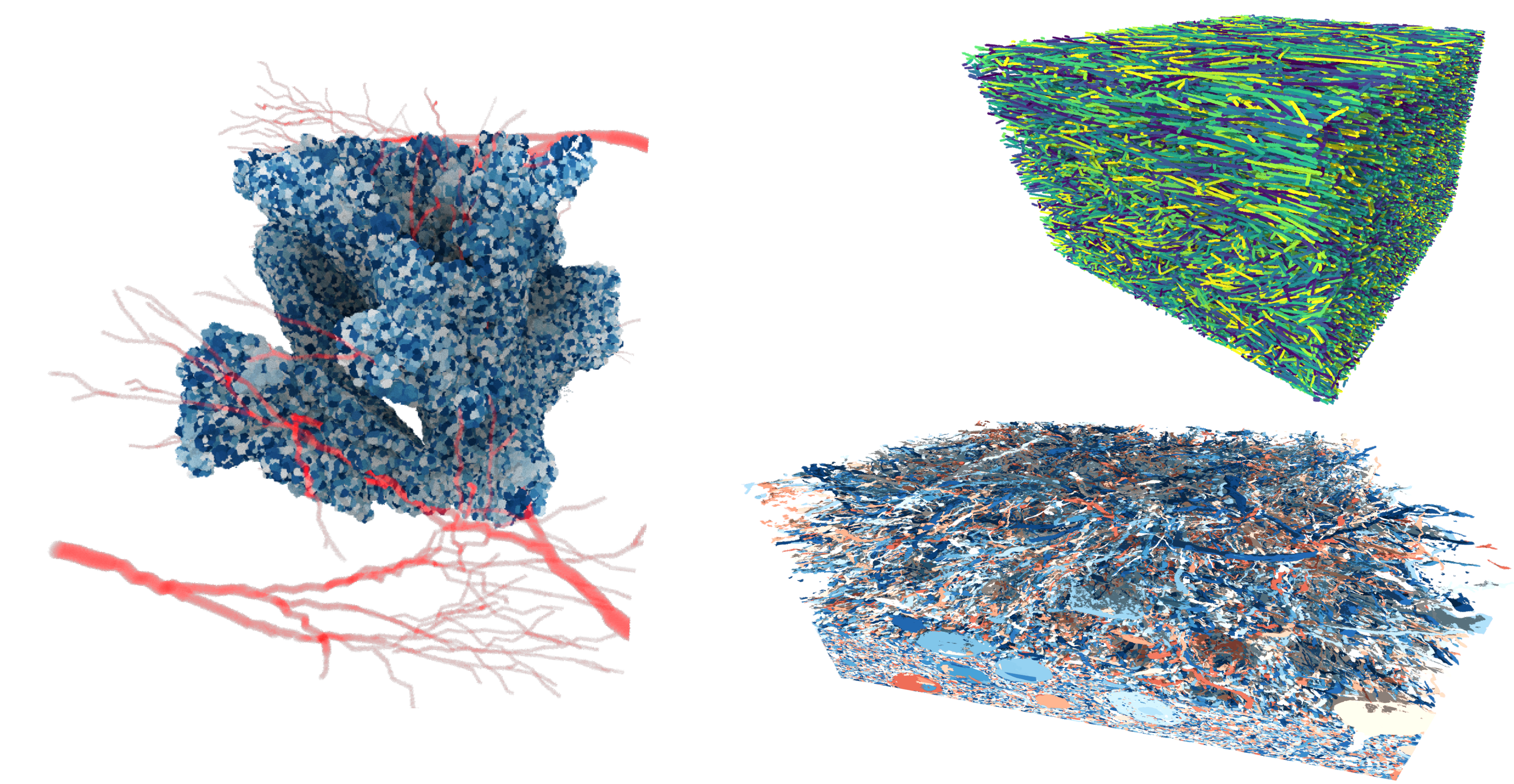}}%
    \small
    \put(0.6,0.4){\makebox(0,0)[ct]{\smash{\begin{tabular}[t]{lr}\textrm{raw 5694 MB }\\\textrm{ours  51 MB  (0.9\%)}\\\textrm{62 fps, 16 ms}\end{tabular}}}}%
    \put(0.6,0.3){\makebox(0,0)[ct]{\smash{\begin{tabular}[t]{lr}\textrm{raw  927.7 GB }\\\textrm{ours   13.5 GB (1.5\%)}\\\textrm{63 fps, 16 ms}\end{tabular}}}}%
    \put(0.01,0.4){\makebox(0,0)[lt]{\smash{\begin{tabular}[t]{lr}\textrm{raw 4000 MB }\\\textrm{ours  112 MB (2.8\%)}\\\textrm{81 fps, 12 ms}\end{tabular}}}}%
  \end{picture}%

  \caption{Our lossless compressed segmentation volumes encode bricks in a multi-resolution fashion followed by entropy coding. Our technique achieves strong compression ratios and provides fast decompression and adaptive level-of-detail. The images show interactive renderings of data sets up to 927 GB uncompressed size (\emph{raw}) at $1920\times 1080$ resolution computed using ray marching with shadows. The raymarching step size is set to match the size of a voxel. Transfer functions (mapping labels to color and opacity) and clipping can be changed interactively.
  }
  \label{fig:teaser}
}

\graphicspath{{figs/}{figures/}{pictures/}{images/}{./}} 

\usepackage{amssymb}
\usepackage{algorithm}
\usepackage[noend]{algpseudocode}
\usepackage{booktabs}  
\usepackage{multirow}  
\usepackage{rotating}  

\usepackage{mathptmx}                  

\begin{document}


\firstsection{Introduction}
\maketitle
In many applications such as in materials science~\cite{Weissenboeck:2014:FS}, connectomics~\cite{Beyer:2022:SVC}, or computational biology~\cite{Berghoff:2020:CIS}, voxel-based segmentation volumes represent how individual regions of interest occupy space in the observed volume by assigning a label to each voxel. 
These volumes can, for example, be obtained from segmenting scalar or multivariate volume data in a preprocessing step~\cite{Motta:2019:DCR}, but they can also be the primary output from simulations~\cite{Berghoff:2020:CIS}.
Segmentation volumes are of fundamental importance when complex structures in large volumes are studied by visual exploration~\cite{Agus:2022:VP}.
However, as with large volume data in general the storage requirements can be challenging, e.g.~when time-series or large ensembles of simulations are computed and stored, or when interactive visualization requires a large portion of a data set to reside in GPU memory for efficient rendering.
Compression techniques can alleviate this problem. In fact a large variety of techniques for volumes storing quantitative data exists~\cite{Rodriguez:2014:SCR}, but such techniques are not directly applicable or beneficial to segmentation volumes as they are meant to represent scalar/vectorial signals. 
In practice, segmentation volumes are often sliced followed by general-purpose image compression techniques. However, this approach is not suited for on-the-fly decompression during visualization and typically less effective than tailored techniques. These, in contrast, exploit the characteristics of segmentation volumes. Compresso~\cite{Matejek:2017:CEC}, for example, encodes the boundaries of segmented regions followed by the Lempel–Ziv–Markov chain compression algorithm (LZMA).
With this combination it achieves high compression ratios, but lacks high decompression speed and the possibility to decompress individual parts of the volume both of which are important for interactive visualization. 

In this paper we propose a novel lossless compression technique for segmented volumes. Our method achieves high compression and decompression speed, strong compression ratios, and is lightweight and well-suited for GPU-based decompression on-the-fly. 
Our compression scheme is based on a per-brick multi-resolution representation of the segmentation volume which exploits the presence of homogeneous label neighborhoods and effectively encodes the boundaries in between. The output of this step is a list of labels and a sequence of operations with which the brick can be reconstructed. For further compression, we store the sequence of operations using a fast asymmetric numeral systems entropy coding scheme (rANS~\cite{Duda:2015:ANS}).
In summary, our contributions are:
\begin{itemize}
    \item a lossless compression technique for segmented volumes with strong compression ratios, little memory overhead, and fast execution times,
    \item a multi-resolution representation for level-of-detail and a sufficiently fine granularity for accessing compressed data,
    \item a parallel, GPU-friendly decompression and caching strategy for interactive visualizations of compressed segmentation volumes, also supporting on-demand streaming of detail information.
\end{itemize}

We will first overview related work on compression for volume rendering. In \cref{sec:encoding} we explain our mulitresolution encoding and decoding scheme for segmented volumes and the subsequent entropy coding. Thereafter we detail the decompression in the context of interactive visualization and GPU-rendering (\cref{sec:rendering}). We evaluate our method on multiple data sets and report compression rates and rendering performance (\cref{sec:results}).

\section{Related Work}
Volume and image compression have been an active area of research for decades~\cite{Hussain:2018:StarIC, Rahman:2019:StarLIC}.
In this overview, we focus on works in the scope of compression of voxel data and segmentation volumes for rendering and scientific visualization.

\paragraph*{Compression of Quantitative Volume Data}
Rodr\'{i}guez et al.~\cite{Rodriguez:2014:SCR} and Beyer et al.~\cite{Beyer:2015:STARLS} provide comprehensive overviews of compression for scalar volumes.
As the input data is often noisy, many existing techniques for scalar volumes perform lossy compression, e.g.~ using wavelets~\cite{Ihm:1999:Wavelet} or neural compression~\cite{Lu:2021:CNR, Weiss:2022:neural}.
As a GPU implementation of the OpenVDB data structure, NanoVDB supports  quantization based compression~\cite{Museth:2021:nanoVDB}.
Lossless techniques exist as well, ranging from run-length encoding~\cite{Rhodes:1985:RLE, Anagnostou:2000:RLE}~\cite{Mados:2019:RLE}, wavelet transforms~\cite{Guthe:2016:LVC}, or Huffman~\cite{Fowler:1994:LCV} and other entropy coders~\cite{Lindstrom:2006:CFD}.
%
In principle, some of these techniques can be applied for segmentation volumes, but since they are tailored to quantitative data they  perform suboptimally.
Still, individual building blocks can be shared, for example, efficient GPU-based implementations for entropy and range coders exist~\cite{Weissenberger:2018:MPH}; we also make use of rANS-coding~\cite{Duda:2015:ANS} in our method. 
Many volume compression techniques make use of hierarchical representations: Sparse voxel octrees~\cite{Laine:2010:SVO} and its  extension to sparse voxel directed acyclic graphs (SVDAGs)~\cite{Kaempe:2013:DAG} are widely used for efficient lossless volume compression in rendering~\cite{Villanueva:2016:SADAG, Careil:2020:IMC, Mados:2020:DAG}.
SVDAGs reuse sub-trees of an initially constructed octree; extensions of the original scheme for binary data can be use to compress arbitrary attributes, making it suitable for a wider range of applications~\cite{Dolonius:2019:CCD}.
Dado et al.~\cite{Dado:2016:GAC} use compressed palettes of voxel attributes that are accessed with an indexing scheme over the graph edges.
Mados et al.~\cite{Mados:2021:SVO} allow replacing homogeneous subregions in DAGs with arbitrary constant values and use variable bit length encoding on voxel attributes stored in leaf nodes.
Note that these methods are optimized for scalar volume data, and they rely on sparsity in the input data to achieve compact representation, a property that segmentation volumes are typically lacking. 

\paragraph*{Segmentation Volume Compression}
Segmentation volumes represent a piece-wise constant, integer-valued function and thus have very different characteristics than the aforementioned volumes. Consequently, specialized compression techniques have been developed.
Neuroglancer~\cite{neuroglancer} splits segmentation volumes into bricks, and each brick is represented by a palette of the contained labels plus one index into the palette for every voxel.
While the approach is fast and well-suited for direct rendering, it does not achieve competitive compression rates.
Compresso~\cite{Matejek:2017:CEC} also uses a brick-wise encoding and determines a set of (binary) templates to represent region boundaries which is then used for encoding. The final strong compression rate is only reached by using a global LZMA compression; however, this makes it unsuitable for GPU-based rendering and decoding of individual bricks.
Our method also compresses individual bricks, however, we avoid a compression of the entire data stream and can inherently decode bricks up to a desired level-of-detail.
The Mixture Graph~\cite{Al-Thelaya:2021:TMG} is a representation of segmentation volumes designed for efficient rendering. It offers a multi-resolution tree hierarchy containing packed label histograms for precise color filtering. 
They use graph compression of histogram factorizations to reduce the memory for their representation. 
While this technique offers a multi-resolution representation, our technique results in an order of magnitude smaller compressed size and faster (de)compression.

\paragraph*{Large-Scale Segmentation Volume Visualization}
Direct rendering of large volume data usually relies on out-of-core methods and streaming~\cite{Beyer:2013:ETC, Beyer:2018:CSV}.
Caching and memory virtualization are used to hide the latency of CPU-to-GPU streaming and to allow direct access of relevant data for the renderer~\cite{Hadwiger:2012:PMS}.
In our exemplary implementation, we also use caching of decompressed volume bricks.
This cache, in particular its implementation on GPUs, is inspired by the Shading Atlas Streaming~\cite{Mueller:2018:SAS} which proposes a texture cache for rendering on untethered virtual reality devices.
Rendering of segmentation volumes is often carried out alongside general volume rendering pipelines~\cite{Choi:2021:ZeVis, Beyer:2013:CE} or within volume segmentation pipelines~\cite{Berger:2018:VAST, Ai-Awami:2016:NB} as well as in systems used for the analysis and information visualization of label region attributes~\cite{Weissenboeck:2014:FS, Troidl:2022:Barrio}.
Agus et al.~\cite{Agus:2022:VP} use dimensionality reduction and topological analysis for guided transfer function design for segmentation volumes.
For large volume rendering, Beyer et al.~\cite{Beyer:2018:CSV} combine probabilistic with exact representations within a hierarchical data structure for fast culling and querying of segmentation data.
Surface extraction and rendering can also serve as an alternative to volume-based rendering of segmentation data~\cite{Lempitsky:2010:SEBV, Newman:2006:SMC, Kobbelt:2001:FSSE}.
In our work, we focus on direct GPU-based rendering of large segmentation volumes. We show that large volumes can be visualized interactively even on modest hardware thanks to strong compression rates and fast decompression, but we also demonstrate CPU-to-GPU streaming of the finest levels of detail for data exceeding the available GPU memory.

\section{Compressed Segmentation Volumes}
\label{sec:encoding}

\begin{figure*}[t]
  \centering
    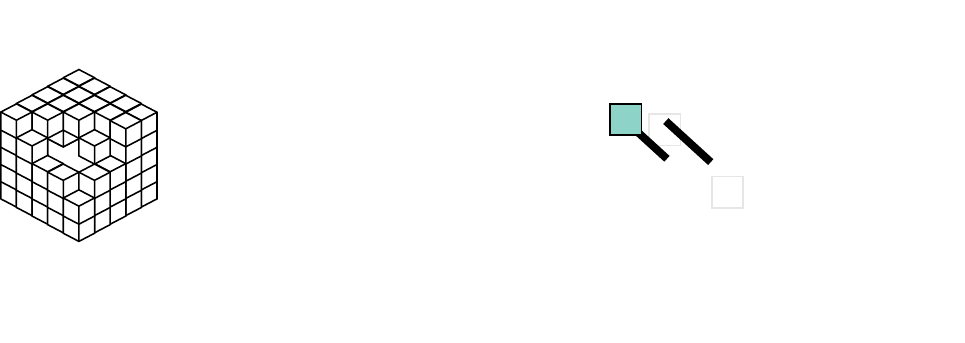
    \caption{A 2D example of the multi-resolution encoding of a brick (different labels shown as colors). 
    First the resolution pyramid with levels $L_N, L_{N-1}, ...$ is built and its grid nodes (starting from the root node, largest/left) are processed coarse-to-fine and along the Morton Z-curve within a level. The traversal order for this brick is shown on the bottom. The result of the encoding is a palette of labels (possibly with duplicates) and a sequence of operations, one for each node. These operations ($P_a$, $R_p$, $R_{\{x,y,z\}}$, $P_0$, and $P_\delta$) define how labels are assigned to nodes, e.g. by reading a label from the palette or reusing the label of its parent or neighbors.}
    \label{fig:codebook_grid}
\end{figure*}

In this section we will describe our compressed segmentation volumes (CSVs). In order to achieve fast and parallel (de)compression with sufficiently fine granularity, our (de)compression operates on volume bricks (Sec.~\ref{sec:SegmentationVolumeBricks}) which are encoded in a multi-resolution fashion (Sec.~\ref{sec:brickencoding}), followed by entropy coding (Sec.~\ref{sec:entropycoding}).

\subsection{Segmentation Volume Bricks}
\label{sec:SegmentationVolumeBricks}
We assume that the input segmentation volumes store a label (an integer value) per voxel and compress individual bricks of the volume separately. We further assume that these bricks all have the same size of $b^3$ labels with $b = 2^N,\ N \in \mathbb{N}$, i.e.~input data might be padded to multiples of $b$ in each dimension.

Our encoding of each brick begins with building a resolution pyramid of $log_2 b + 1$ levels. We denote the finest level storing $b^3$ labels as $L_0$, and successively compute the coarser levels $L_l$, $0<l\leq N = log_2 b$. In each step the resolution is halved along each axis, i.e.~$L_l$ stores $(b/2^l)^3$ labels, and each voxel in $L_l$ overlaps $2^3$ voxels in $L_{l-1}$. For the label of a voxel in $L_l$ we assign the most frequent label in the corresponding 8 voxels in $L_{l-1}$. 
As this resolution pyramid essentially forms an octree, we will refer to voxels in the pyramid as \emph{nodes} when describing the encoding and decoding (see \cref{fig:codebook_grid}). 
We generate the pyramid explicitly for a brick during encoding which requires about 14\% additional temporary memory. The encoded representation implicitly contains the multi-resolution representation; decoding does not require more memory than necessary to store the $(b/2^l)^3$ labels of the desired level-of-detail. 

\subsection{Brick Encoding}
\label{sec:brickencoding}

The key to a good compression is to compactly encode the assignment of labels to voxels. Our encoding can leverage local homogeneity, e.g.~by copying labels from parent or neighbor nodes, and only rarely reading new labels from a separate list which we call the \emph{palette}.

\begin{figure}[tbp]
  \centering
    \def\svgwidth{0.6\columnwidth}
    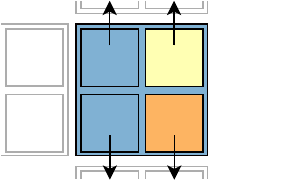
    \caption{The operations $R_{\{x,y,z\}}$ assign the same label to the currently processed node as found in an adjacent nodes. They define the axis along which this neighbor is found, the direction always points outside the current $2^3$ block of voxels ($2^2$ in this 2D-example) with the same parent node.}
    \label{fig:neighbor}
\end{figure}

Our encoding, and likewise later the decoding, begins with the coarsest level $L_N$, which represents the root node of the octree. The label of this node is the first label stored in the palette; we denote the index of the last used palette entry as $i_p$ which is initialized to zero. The respective child nodes in $L_{N-1}, L_{N-2}, ...$ then need to be processed in a defined order, for which we use a Morton Z-curve in each level; a concatenation of the Z-curves yields the enumeration of all octree nodes (\cref{fig:codebook_grid}).
The core idea of our encoding is that the label of the next node in this order often can be determined by a simple operation, such as assigning the same label as the parent node, or reading the next label stored in the palette; the result of a brick encoding then becomes a sequence of operations and the associated palette.
In the following, we introduce and discuss the individual operations we have chosen. The selection is a result of investigating typical configurations in the resolution pyramids and compression experiments; it comprises the following operations:
\begin{itemize}

    \item {\bf Parent reuse} $R_p$: assign the label of the (coarser) parent node to the next node. Note that the processing order of nodes guarantees that the parent node's label is known. For parent nodes we chose the most frequent label among its children, consequently this operation is often applicable.

    \item {\bf Palette Advance} $P_a$: increase $i_p$, read the label at the new index from the palette, and assign it to the next node.

    \item {\bf Neighbor Reuse} $R_x$, $R_y$, $R_z$: these operations are used to reference nodes adjacent to the next node and assign their label. Note that we reference nodes \emph{outside} the $2^3$ block of sibling nodes only (see \cref{fig:neighbor}). This operation can  refine boundaries by ``pulling in'' the label from neighboring regions. 
    Thus the operations $R_{\{x,y,z\}}$ only define along which axis we reference while the direction is implicit.
    If the referenced node has a later Z-index, however, it is not yet decoded. In this case, we reinterpret a neighbor reuse as a reference to the neighbor's parent whose label is known. Neighbor nodes in other bricks are not referenced.
    \cref{fig:codebook_grid} illustrates such an operation used to define the shape of the yellow region.

    \item {\bf Palette Back References} $P_{0}$, $P_\delta$: 
    When reuse-operations are not applicable, palette advance operations (which require storing another label) can be avoided by back-referencing a previously used palette entry. $P_\delta$ references the palette entry at index $i_p-\delta-1$ ($\delta \in [0, 15]$ is stored as 4-bit value), and the special case operation $P_{0}$ indexes the last used palette entry $i_p$; the respective entry is assigned to the next node.

\end{itemize}

The children and children's children of nodes on finer levels often represent the interior of homogeneous regions and thus carry the same label.
To not store redundant operations, we output one additional \emph{stop bit} for every node to indicate whether or not further operations follow for the respective nodes in finer levels.

\paragraph{Discussion}
In our experiments we considered additional operations, such as reusing diagonal neighbors or grandparents, but found that small operation sets typically yield better results.
We also considered resolving references $R_{\{x,y,z\}}$ to not yet decoded neighbors, but this leads to possibly long chains of dependencies.
Given that referenced neighbors always have a different parent than the processed node and that those parents often carry the correct label this does not noticeably impact the compression. Lastly, note that palettes can still contain duplicates, e.g. when a label has previously been used in the decoding of a brick, but cannot be referenced with $P_\delta$.

Our encoding operates along the Morton Z-curve, and only $P_0$ and $P_\delta$ refer to previous nodes along the curve.
However, the traversal order determines which neighbor references by $R_{\{x,y,z\}}$ have already been en/decoded. As an alternative we tried Hilbert curves which led to almost identical compression rates. While each (non-boundary) node can still reference 3 neighbors on the current level \emph{on average}, their number and the directions to valid references depend on the position along the Hilbert curve (in contrast to Morton Z-curves). The more costly evaluation of the Hilbert curve and the additional cost for referencing result in 2 to 3 times longer compression times.
The supplemental material contains a discussion in more detail.

\begin{algorithm}[t]
\caption{Decoding a brick up to a target level-of-detail $t$.}
\label{alg:decoding}

\hspace*{\algorithmicindent} \textbf{Input} \hskip 1em brick \textit{operation stream} and \textit{palette}, target LOD \textit{t}\\
\hspace*{\algorithmicindent} \textbf{Output} \hskip 0.3em array \textit{out} containing decompressed brick in LOD \textit{t}
\begin{algorithmic}[1]

\State $i_p \gets 0$ \Comment{palette read index}
\State $out[0 \dots 2^{3(N-t)}] \gets \{\varnothing \dots \varnothing\}$ \Comment{initialize output array}
\State $out[0] = palette[i_p]$

\vspace{0.2cm}
\For{$l \in [N \dots (t+1)]$} \Comment{coarse to fine}

\ForAll{already decoded node (spacing $2^{l-t}$) in Z-order}

    \State $i \gets$ index of the node in $out$
    \State {\bf if} {$out[lastChildOf(i)] \neq \varnothing$} {\bf then continue } \Comment{constant area}

    \vspace{0.1cm}
    \State $parent \gets out[i]$  \Comment{store parent for next 8 nodes}
    
    \vspace{0.1cm}
    \ForAll{child node (spacing $2^{l-t-1}$) in Z-order}
        \State $j \gets$ index of the child node in $out$
        \State $(op, \ stop) \gets readNextOperationAndStopBit()$
        \vspace{0.2cm}
        \State {\bf switch($op$)}
        \State \hspace{1em} {\bf case} $R_p$ : $out[j] \gets parent$
        \State \hspace{1em} {\bf case} $R_{\{x,y,z\}}$ : $out[j] \gets neighbor\ value$
        \State \hspace{1em} {\bf case} $P_a$ : $out[j] \gets palette[ ++ i_p ]$
        \State \hspace{1em} {\bf case} $P_0$ : $out[j] \gets palette[ i_p ]$
        \State \hspace{1em} {\bf case} $P_{\delta}$ : $out[j] \gets  palette[ i_p - \delta ]$
    
        \vspace{0.2cm}
        \If {stop}
            \State fill $j$'s entire sub-block with $2^{3(l-t-1)}$ labels in $out$
       \EndIf
    \EndFor
\EndFor
\EndFor
\end{algorithmic}
\end{algorithm}
\begin{algorithm}[t]
\caption{Encoding a brick into a palette and sequence of operations.}
\label{alg:encoding}
\hspace*{\algorithmicindent} \textbf{Input} \hskip 1em original \textit{brick} voxels from volume\\
\hspace*{\algorithmicindent} \textbf{Output} \hskip 0.3em brick \textit{operation stream} and \textit{palette}
\begin{algorithmic}[1]
\vspace{1.4em}

\State $pyramid \gets $ brick's multi-resolution pyramid to encode
\State $palette \gets ${label of pyramid's root node}

\vspace{0.2cm}
\For {$l \in [N \dots 1]$}

\ForAll {node on level $l$ (spacing $2^{l}$) in Z-order}

    \State $i \gets$ index of the current node
    \State {\bf if} {$pyramid[i].$constantChildren} {\bf then continue }

    \vspace{0.1cm}
    \State $parent \gets pyramid[i]$.label  \Comment{parent for next 8 nodes}
    
    \vspace{0.1cm}
    \ForAll{child node (spacing $2^{l-1}$) in Z-order}
        \State $j \gets$ index of current child node

        \vspace{0.2cm}
        \State $L \gets$ $pyramid[j].$label
        \State $stop \gets$ $pyramid[j].$constantChildren

        \vspace{0.2cm}
        \State $op \gets bestOperation(parent, pyramid, palette, L)$
        
        \vspace{1.825em}
        \State {\bf if} {$op = P_a$} {\bf then} $palette$.push($L$)
        \State {\bf if} {$op = P_\delta$} {\bf then} output $\delta$
        \vspace{1.825em}
        \State output (op, \ stop)

    \EndFor
\EndFor
\EndFor
\end{algorithmic}
\end{algorithm}

\subsection{Entropy Coding of Operation Sequences}
\label{sec:entropycoding}
The representation of a brick as a sequence of operations already reduces storage, but each operation occupies at least 4 bits (7 different operations plus stop bit; in case of $P_{\delta}$, 4 additional bits for $\delta$). As the frequencies of operations are highly imbalanced (see \cref{fig:operation_frequency}) we apply an entropy coding to further reduce storage. We found that using range asymmetric numeral systems (rANS)~\cite{Duda:2015:ANS} is a good compromise between Huffman coding (fast, but suboptimal because of the fixed number of bits per symbol) and arithmetic coding (slower).
We directly use the sequence of 4-bit nibbles as data stream. Interestingly, the frequencies of these 
nibbles are extremely similar across the bricks of an entire segmentation volume (see \cref{sec:res_frequency} for details). This enables us to quickly determine static, well-suited frequency tables for a data set, and also later efficiently perform the rANS-decompression and execution of operations in one go and in parallel for the volume's bricks.
Note that we create two frequency tables per data set: one for the interior nodes (levels $L_{n} .. L_{1}$), and one for leaf nodes (level $L_0$) whose stop bits are always 0.
Of course parallel (de)compression is also possible with individual frequency tables per brick or adaptive frequencies, but this requires additional storage for the tables or results in worse compression ratios when frequencies need to adapt to the data stream first.

\subsection{Encoding and Decoding Implementation}
In this section we discuss important algorithmic details and begin with the decoding of a brick from a given palette and sequence of operations and stop bits. 

\begin{figure}[tbp]
  \centering
    \includegraphics[width=0.8\columnwidth]{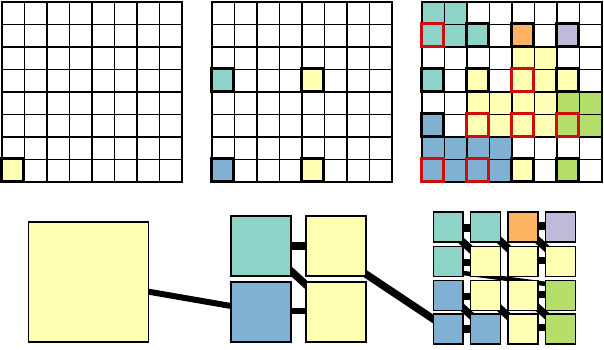}
    \caption{This example shows three decoding steps $L_{N}, ..., L_{N-2}$ of a brick (in 2D).
    The grids (top) show the in-place decoding and are sized to store the brick at target LOD $t=N-3$, i.e.~$2^{3(N-t)}$ labels ($64=2^{2(N-t)}$ in 2D).
    The nodes of the resolution pyramid and the Z-curve order is shown on the bottom. 
    Nodes highlighted in red have a stop bit set and assign their label to all their child nodes; for these no further decoding operations are required.}
    \label{fig:output_block}
\end{figure}

\paragraph{Decoding}
The decoding begins with the coarsest level $L_N$ and can be performed up to a desired target level-of-detail $L_t, t=N..0$. This eventually results in a block of ${2^{3(N-t)}}$ labels. Prior to decoding, the memory for this 3D-array output is allocated and the decoding is then performed in-place: While processing a level $L_l, l=N..t$, its labels are stored with a spacing of $2^{t-l}$ in the output array; all entries are filled when the decoding is complete. \cref{fig:output_block} shows a 2D-example with three intermediate decoding steps. 
\Cref{alg:decoding} details the decoding procedure of a single brick. After initialization (lines 1-3) it processes one level-of-detail after another, beginning with the coarsest level $L_N$ (line 4). Line 5 loops over the $2^{3(N-l)}$ nodes on level $l$ in the Morton Z-curve order. If a label has already been assigned to all of its child nodes (line 7), this is because a stop bit has been set on a coarser level (lines 18-19) and no further decoding for this node's children is required\footnote{This test is performed redundantly down to leaf nodes. However, homogeneous regions are typically not large, i.e.~stop bits are typically set at finer levels and the overhead remains small. Still the use of stop bits avoids storing superfluous operations for many nodes.}.
Otherwise, the node's label is temporarily stored in \emph{parent} (line 8), and its eight child nodes are decoded by reading the next 4-bit tuple of operation and stop bit (line 11). The operation determines the label of the next child node (lines 12-17). If the stop bit is set, the entire sub-block of $2^{3(l-t-1)}$ labels is set (lines 18-19).
Note that we overwrite the previously decoded coarser LODs until we reach $t$, but at the expense of higher memory consumption overwriting is not mandatory per se.

\paragraph{Encoding}
Similar to decoding, all bricks can be encoded in parallel; the pseudo-code is given by \cref{alg:encoding}.
The encoding of a brick begins with computing the resolution pyramid ($pyramid$) which requires about 14\% additional temporary storage for the brick (line 1), similarly to a 3D mipmap~\cite{Williams:1983:PP}.
The pyramid contains each node's reference label, and if the node's subtree is constant, the stop bit is set (line 11).
If there is an ambiguity when determining the most frequent child label for a pyramid node, we chose the one occurring first within the child node array.
The encoding is performed analogous to the decoding, but of course it determines which operation is suitable to yield the label of each child node (line 12).
For this, \textit{bestOperation} tests for the first possible operation in a fixed order which we found a good fit to their typical frequencies: $R_p$, $R_x$, $R_y$, $R_z$, $P_0$, $P_\delta$, $P_a$.
Sticking to this fixed order also additionally skews the operations' frequencies favorably for entropy coding.
For any $P_\delta$ we directly output $\delta$ as an additional 4-bit entry to the encoding stream (line 14).

Note that when compressing a segmentation volume, we typically feed the output nibbles of a brick directly into the entropy coder. For this, prior to the actual encoding, we perform a quick prepass over a subset of bricks (in our examples every 512$^{th}$ brick, or every 4096$^{th}$ for large data sets) to determine the two frequency tables of nibbles for both interior and leaf nodes. Using two frequency tables has no performance impact on the rANS-encoding, and the additional storage is negligible.

\section{Visualization of Compressed Segmentation Volumes}
\label{sec:rendering}

\begin{figure}[t]
  \centering
    \includegraphics[width=0.9\columnwidth]{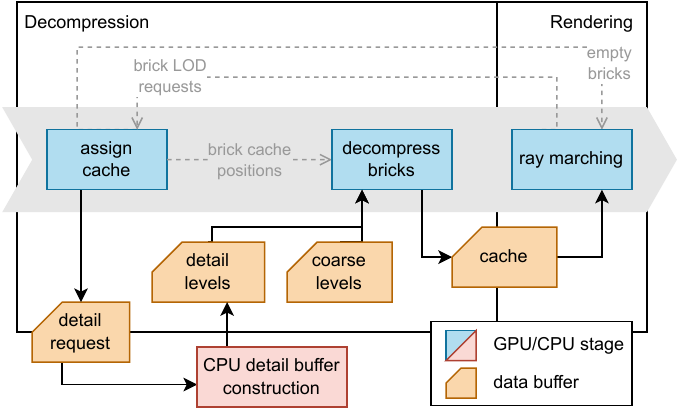}
    \caption{An overview of our renderer: Using raymarching we read voxel-data from bricks stored in a cache. For intersected bricks that are not present in the cache, or not at the required level of detail, we generate a request. These bricks are assigned and decompressed to a free cache location, and can be accessed in the next frame. If possible, the compressed data completely resides in GPU memory. If a CSV is too large, the finest levels of detail are streamed from CPU memory to the GPU on demand.}
    \label{fig:pipeline}
\end{figure}

In this section we describe how compressed segmentation volumes (CSVs) can be used with raymarching for volume visualization. We assume that transfer functions (TFs) are used which map labels to color and opacity; note that the design of TFs is orthogonal to our work.
Our exemplary raymarcher works similar to other bricked volume visualizations and efficiently supports empty-space skipping of bricks that are invisible due to the TF. Bricks are decompressed with the required level-of-detail (LOD) on demand and then stored in a cache in order to facilitate fast accesses during raymarching and to exploit temporal coherency in the camera movement (and thus in visible bricks and LODs).
For an overview of basic raymarching and volume shading techniques we refer the reader to J\"{o}nsson et al.~\cite{Joensson:2014:SVI}. We detail the technical details of our implementation in Sections~\ref{sec:vis_caching} and ~\ref{sec:detail_separation} (see also Fig.~\ref{fig:pipeline}). The next paragraphs provide a high level summary of the most important aspects.

\paragraph{Raymarching} We perform straightforward raymarching using one ray per pixel (\cref{fig:raymarch}). Invisible bricks are skipped (see below). For visible bricks intersected by a ray we determine their required LOD. For our tests, we choose the LOD based on the distance of a brick's center to the camera such that one voxel maps to approximately one pixel on the screen. If the desired LOD is not available in the cache, it is requested and it will be decompressed for the next frame (purple, \cref{fig:raymarch}).
We allow this lag of one frame as the resulting artifacts are typically small when the frame rate is reasonably high. This also allows us to focus our tests on the compression scheme as the core of our method, but of course more elaborate schemes to access the desired LOD within the same frame can be used (e.g. akin to~\cite{Hadwiger:2018:SparseLeap}). Note that rendering can always fall back to the coarsest LOD of a brick if it has not yet been decompressed, as this level is stored as the first palette entry and thus directly accessible.
Bricks are flagged upon sampling during raymarching so that bricks that become invisible can be deallocated (red, \cref{fig:raymarch}) in the next frame.
To that end, we reset a buffer of one integer per brick to an \textit{invisible} flag before each frame and let all ray marching threads write requested LODs in parallel. Since a requested LOD depends on the brick's center only, these are identical between threads and no race conditions occur.
For hit voxels above a user-defined opacity threshold we send an additional shadow ray. Optionally we can approximate ambient occlusion by accumulation shadow rays over time (see \cref{fig:teaser}, left).

\paragraph{Empty-Space Skipping} Recall that for every compressed brick we store the palette of labels; this palette has orders of magnitude fewer entries than there are voxels in the brick. If none of the palette entries is mapped to an opacity greater than 0, then the entire brick can be skipped during raymarching and also does not need to be decompressed as long as the TF does not change its visibility.

\paragraph{Decompression and Caching} The decompression is executed in parallel for all required bricks and their LODs directly on the GPU, and the result is put into a cache in GPU memory. 
The CSVs typically also reside in GPU memory, but in order to be able to render very large CSVs we optionally keep the compressed data for decoding the finest level(s) of detail in main memory and transfer it to the GPU on demand. 
Note that the fine LODs consume a significant portion of the data, but are only required for regions of the CSV which are visible and close to the camera.

\begin{figure}[t]
  \centering
    \includegraphics[width=0.9\columnwidth]{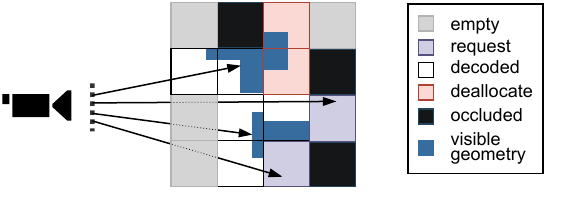}    
    \caption{Our raymarching is kept simple to focus the evaluation on the decompression performance: We march along one ray per pixel (plus one shadow ray) and perform empty-space skipping on a brick level. When the raymarcher tries to access a brick that is not present in the cache, or present but at the wrong level of detail, the brick is requested for the next frame. Bricks that are not accessed in the current frame are simply evicted from the cache.}
    \label{fig:raymarch}
\end{figure}

\subsection{Brick Caching and Decoding}
\label{sec:vis_caching}
\begin{figure*}[tbp]
  \centering
    \includegraphics[width=\textwidth]{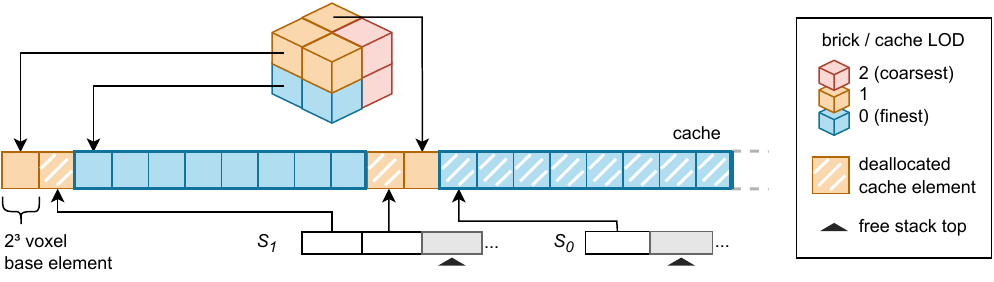}
    \caption{
    Our cache is organized as base elements of $2^3$ voxels which are combined to form larger blocks. In this example, bricks of LOD 1 correspond to 1 base element ($2^3$ voxels), LOD 0-bricks to 8 base elements ($4^3$ voxels). To quickly assign free locations to bricks during decompression, we use one stack $S_l$ per LOD where one stack element points to the first base element where a block of the respective size can be stored. The locations of bricks evicted from the cache are pushed onto these stacks.
    Note that the coarsest LOD (here level 2) is never stored in the cache as it is available as the first entry of a brick's palette.
    }
    \label{fig:cache}
\end{figure*}
Our cache is loosely inspired by Shading Atlas Streaming~\cite{Mueller:2018:SAS} (SAS), which has been designed to stream 2D texture tiles from a powerful host to a tethered VR device.
We adapt a subset of SAS to cache decompressed CSV bricks on a single GPU. 
In particular, we adopt the handling of free cache elements with one stack per LOD from SAS, and their usage of atomic counters to assign bricks to stack elements.
However, as our LOD scheme does not create anisotropic bricks, we omit the superblock and column concepts from SAS.
Instead, we simply map contiguous base elements to bricks.

For this our cache maintains a pool of base elements of $2^3$ voxels which are allocated and combined to form \emph{blocks} of sizes $2^{3l}, l=1..N$ which can then store one decompressed brick of the respective size (\cref{fig:cache}).
Bricks at the coarsest resolution $l=0$ do not need to be decompressed.
As in SAS, the index of the first base element of an already formed but currently unused block is stored on stacks $S_l$ (one stack per block size) in order to quickly retrieve free blocks when decompressing bricks. If a stack $S_l$ runs empty, the cache allocates new blocks of size $2^{3l}$ by atomically advancing a shared cache top pointer by $l$. After a block has been assigned to a level $l$, it cannot be used for other levels later. If no free base elements are available, we resort to a rebuild of the cache as proposed by SAS to remove fragmentation. Note that for reasonably sized cache pools this happens extremely seldom.

The use of the cache during visualization is as follows: 
After rendering a frame, the required bricks and their LODs for the next frame are known. They will then be decompressed and stored into the cache (\emph{assign cache} in \cref{fig:pipeline}).
For this, we first test whether a requested brick is visible for the current TF, and if visible, we mark this brick by assigning an ascending index (one sequence per block size $2^l$ as proposed by SAS) to it. 
By this, we also count how many bricks of which LOD need to be decompressed.
Bricks already in the cache that became invisible because of the TF or not being hit by a ray are evicted from the cache and their blocks are pushed onto the respective stack $S_l$ (\cref{fig:cache}).

Next we determine for each $S_l$ if it contains enough free blocks to fulfill all requests of level $l$.
If this is not the case, the cache allocates the new blocks required to decompress all requested bricks.
As mentioned before, if not enough blocks can be allocated, the cache is rebuild, i.e.~all blocks and stacks are cleared and all visible bricks of the CSV are newly decompressed.
Lastly, the actual decompression of each requested brick (\emph{decompress bricks} in \cref{fig:pipeline}) is carried out in parallel.

The architecture of our cache follows that of SAS with the following differences: 1) we test bricks for empty space using their palettes before requesting free locations, 2) we avoid superblocks by providing contiguous base elements using the method from above, and 3) during assignment of locations we decompress the bricks.

\subsection{Detail Separation}
\label{sec:detail_separation}
If possible we store the CSVs in GPU memory. For too large CSVs we split the compressed data: the palette and data for the coarse levels of detail are kept on the GPU, and data for further decompressing fine detail levels is stored in CPU memory.
While the latter needs a comparatively large amount of memory, bricks with high (or full) detail are often only required for close up views. The \cortex{} data set (see \cref{fig:teaser}), for example, has a resolution of $6144\times 9216\times 4096$ voxels, i.e. even at high rendering resolutions only the front most bricks might be needed with high detail. On the other hand, coarser levels are frequently accessed, also during decompressing finer levels.

To this end, we stream the detail levels for bricks to the GPU only when required (\cref{fig:pipeline}, bottom).
{For this, the cache assignment stage (Section~\ref{sec:vis_caching}) determines for every brick if detail levels need to be accessed for decompression.}
In this case, the brick index is added to a \emph{request buffer} which is transferred to CPU memory after the assignment. A buffer containing the requested detail levels is generated in parallel to rendering the next frame and asynchronously uploaded to GPU memory.
We implemented two options to handle this resulting additional lag of one frame:
first we can simply decompress a brick only up to the finest available level in GPU memory, or second, we predictively upload detail information for bricks close to the camera.
Once a brick has been decompressed and stored in the cache, the data for fine details is not required anymore and can be discarded.
Note that in practice, even for the largest data sets tested we were able to store all but the data for decompressing level $L_0$ in GPU memory.
For our compression of \cortex{}, the total memory consumption for $L_0$ is 9.2 GB while all other levels take up only 4.3 GB.
In our experiments we use an 8 MB buffer to upload requested detail level data to the GPU which was sufficient in most of the frames. 
If more detail data is requested (or the buffer would be smaller) the upload is distributed across several frames trading LOD adaptation for responsivity/interactivity.

\begin{table*}[t]
\centering
\caption{Compression rates and CPU compression times of different data sets and brick sizes, without and with using rANS-entropy coding. The third column shows the figures for our default: using rANS with two global frequency tables, one for $L_N .. L_1$ and one for $L_0$. Timings are measured excluding data set input/output operations. For larger brick sizes we use 8 instead of 16 threads (marked by $()^+$ after the timing value).}

\label{tab:compression}
\begin{tabular}{cc|lll|lll|lll}
&  & \multicolumn{3}{|l}{\bf no rANS} & \multicolumn{3}{|l}{\bf  rANS, one frequency table} & \multicolumn{3}{|l}{\bf  rANS, two frequency tables}\\
& b & CR & Time (s) & GB/s & CR & Time (s) & GB/s & CR & Time (s) & GB/s\\
\midrule
\multirow{3}{*}{\rotatebox{90}{\cells{}}} & $16$ & 6.958\% & 2.008 & 1.992 & 3.980\% & 2.104 & 1.901 & 3.561\% & 2.090 & 1.914\\
                                        & $32$ & 6.581\% & $1.859^+$ & 2.152 & 3.460\% & 1.913 & 2.022 & 3.016\% & 1.957 & 2.043\\
                                        & $64$ & 6.428\% & $2.000^+$ & 2.000 & 3.250\% & $2.189^+$ & 1.827 & {\bf 2.805\%} & $2.161^+$ & 1.851\\
\midrule
\multirow{3}{*}{\rotatebox{90}{\fiber{}}} & $16$ & 3.220\% & 3.579 & 1.591 & 1.950\% & 3.812 & 1.494 & 1.737\% & 3.887 & 1.465\\
                                        & $32$ & 2.597\% & 3.017 & 1.887 & 1.257\% & 3.148 & 1.809 & 1.017\% & 3.744 & 1.521\\
                                        & $64$ & 2.502\% & $3.457^+$ & 1.647 & 1.147\% & $3.601^+$ & 1.581 & {\bf 0.899\%} & $3.219^+$ & 1.769\\
\midrule
\multirow{3}{*}{\rotatebox{90}{\cortex{}}} & $16$ & 3.684\% & 284.475 & 3.261 & 2.035\% & 360.136 & 2.575 & 1.882\% & 362.248 & 2.561 \\
                                         & $32$ & 3.357\% & $294.042^+$ & 3.155 & 1.699\% & $328.956^+$ & 2.820 & 1.533\% & 320.558 & 2.894 \\
                                         & $64$ & 3.307\% & $321.660^+$ & 2.884 & 1.631\% & $351.806^+$ & 2.637 & {\bf 1.459\%} & $347.999^+$ & 2.666\\
\end{tabular}
\hfill
\begin{tabular}{l}
\\
data set characteristics\\
\midrule
$1000\times1000\times1000$ voxels\\
1,067,198 labels\\
4.0 GB uncompressed\\
\midrule
$1579\times1092\times1651$ voxels\\
31,877 labels\\
5.7GB uncompressed\\
\midrule
$6144\times9216\times4096$ voxels\\
15,030,572 labels\\
927.7GB uncompressed\\
\end{tabular}
\end{table*}%

\subsection{Possible Optimizations}
\label{sec:optimizations}
Here we briefly mention possible improvements for future work. There are plenty of performance optimizations known for large-volume visualization which can be combined with CSVs. For example, empty space skipping could use an octree where leaf nodes represent CSV-bricks to efficiently detect larger invisible regions during raymarching.
It is also obvious that the caching strategy can be refined, e.g.~by predicting bricks becoming visible within the next frames when they move towards the view frustum. Noteworthy is also that there is a large body of work on occlusion culling, either view-dependent or computed globally~\cite{Ernst:2004:Entkerner}, which can be used to reduce cache usage and increase rendering performance.

\section{Results}
\label{sec:results}

In this section we evaluate our CSVs, compare our compression to previous work (Section~\ref{sec:compressionperformance}, and discuss the rendering performance in Section~\ref{sec:renderingperformance}). 
We will make the source code of our compression technique available.
For the evaluation we use the following segmentation volumes (see \cref{fig:teaser}) taken from simulations and measurements:
\begin{itemize}
    \item 
\cells{}: A Cellular Potts Model cancer growth simulation~\cite{Rosenbauer:2020:ETD} with a resolution of $1000\times1000\times1000$ voxels and 1,067,198 labels (1067 labels/million voxels).
As each individual biological cell in such simulations has its own label, this data set has by far the most labels per voxel in our evaluation.

    \item 
\fiber{}: A fiber segmentation of an X-ray scan of a glass fiber reinforced polymer~\cite{Maurer:2022:Fiber} with $1579\times1092\times1651$ voxels and 31877 labels (11 labels/million voxels).
Due to the low number of labels, this data set has been provided with 16 bit per voxel as opposed to the other 32 bit data sets in this evaluation. 
The segmentation volume contains highly anisotropic label regions and inhomogeneously shaped empty space that leads to a partial visibility for many bricks during rendering. 

    \item 
\cortex{}: A segmentation of an electron microscopy scan of a mouse cortex~\cite{Motta:2019:DCR} with $6144\times9216\times4096$ voxels and 15,030,572 labels (65 labels/million voxels).
This is the largest data set is in our evaluation with an uncompressed size of almost 928 GB and label regions of strongly varying size and shape. The resulting CSV requires storing detail level data for $L_0$ on the CPU. If a brick is required in full detail and $L_0$ has not yet been uploaded to GPU memory, the decompression temporarily uses $L_1$ (see Section~\ref{sec:detail_separation}).
\end{itemize}

\subsection{Compression Performance}
\label{sec:compressionperformance}

\begin{table*}[t]
\centering
\caption{Compression rates of different techniques: hdf5 uses gzip on bricks of $128^3$ voxels, for png we sliced the volume and used zlib with the highest compression level. For Compresso we used the default parameters with a window size of (8,8,1). For Neuroglancer we set the block size to (8,8,8). Our technique (with $b=64$) is benchmarked without and with rANS using two frequency tables to assess the overhead of entropy coding. Timings are reported for single-threaded execution; the timings for our method running with 16 threads are shown in parentheses.
Note that the full \cortex{} data set did not fit into memory and the available implementations of Compresso and Neuroglancer did not handle out-of-core compression. To this end, we chose a representative $1024^3$ subvolume of \cortex{} which matches the overall average number of labels per voxel and whose hdf5-compression ratio is  the same as for the entire \cortex{} data.}
\label{tab:comparison}
\begin{tabular}{ll|r|r|r|r|r|r|r}
data set & size/\#labels & hdf5 & png & Compresso & Compresso+LZMA & Neuroglancer & ours (no rANS) & ours (rANS) \\
\midrule
\cells{}    & 4.0GB & 7.221\% & 10.812\% & 8.337\% & 2.753\% & 13.622\%  & 6.428\% & 2.805\% \\
            & 1M labels &  23s    & 493s     &   35s   & 135s    &    11s    & 15s (3s) & 16s (3s)\\           
\rule{0pt}{2ex} \fiber{} & 5.7GB      & 3.051\% & 3.665\% & 26.700\% & 5.861\% & 3.658\% & 2.502\% & 0.899\% \\
                         & 32K labels &  53s    &    380s &   130s &  619s & 12s & 19s (6s) & 20s (6s)\\
\rule{0pt}{2ex} \cortex{}$^*$ & 4.3GB      & 2.459\% & 2.406\% & 8.515\% & 1.267\% & 3.999\% & 3.564\% & 1.590\%\\
                              & 45K labels & 18s     &    138s & 35s     & 138s    & 5s      & 10s (3s)  & 11s (3s)\\
\end{tabular}
\end{table*}

In this section we evaluate the compression rate for a variety of settings and compare our method to previous work. We measure time and throughput for the compression and define the reported compression rate (CR) as the ratio of compressed size to input size, i.e.~smaller values mean better compression.

\paragraph{Parameters for CSV-compression}
We measure all timings on an AMD Ryzen 7 5800x 8-core CPU with 64 GB of RAM.
Data sets that do not fit into RAM at once, e.g. the \cortex{}, can trivially be processed in an out-of-core fashion as the (de)compression is performed for individual bricks.
{Note that we used a naive parallelization throughout our experiments: $T$ threads simply compress $T$ bricks in parallel, followed by a synchronization and concatenation of the output. In particular for larger brick sizes, this results in less optimal utilization as the processing times between threads diverge. We leave more elaborate schemes, e.g.~using work queues, as well as a GPU implementation for future work.
}

\Cref{tab:compression} summarizes the results for the data sets and shows results for different brick sizes $b$ as well as compression with and without rANS entropy coding; we use 16 threads for smaller and 8 threads for larger brick sizes.
The results show that larger brick sizes (with proportionally fewer edge voxels) lead to better compression rates.
This is mainly because neighbor references $R_{\{x,y,z\}}$ cannot be used across the bricks' borders. Depending on the data set, large homogeneous regions that span multiple bricks are more compactly encoded by with fewer large bricks.
The rANS coding improves the overall CRs by about 40\% compared to storing 4 bit nibbles directly.
As mentioned before, for the compression with rANS we compute two global frequency tables per data set---one for nodes in $L_0$ where the stop bit is always $0$, and one for all other nodes---during an encoding prepass using every $512^{th}$ brick, except for \cortex{} where we use every $4096^{th}$ brick (Section~\ref{sec:brickencoding}).  
In our experiments, CRs only varied very subtly when increasing or decreasing the subsampling.
We observe compression speeds of 1.5 to 3 GB/s for all data sets; compression with $b=32$ consistently achieves the highest throughput, despite the lower utilization of CPU cores.
rANS coding only slightly impact the throughput by about 2.5\% on average.

\begin{figure}[tbp]
  \centering
    \includegraphics[width=\linewidth]{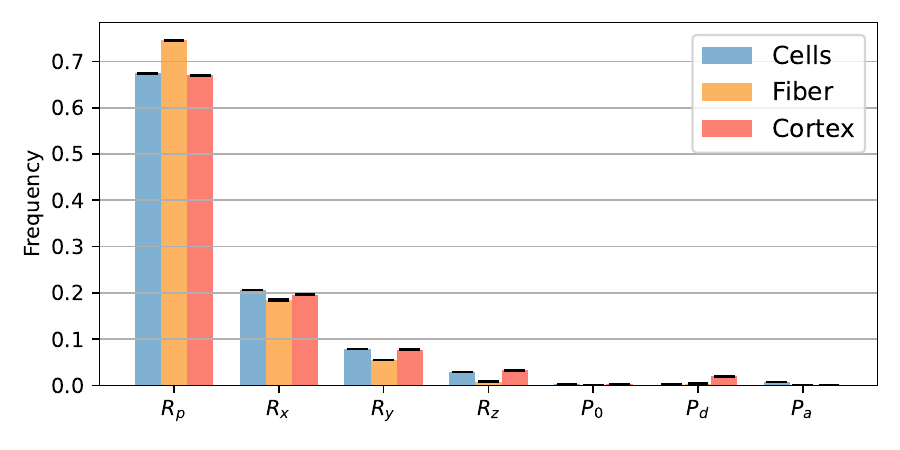}
    \caption{Mean relative frequencies of operations for the \cells{}, \fiber{}, and \cortex{} data sets.
    We also show the standard deviation $\sigma$. Note that we did not compute it from the bricks' operation counts directly, as some bricks can be encoded with very few operations and would lead to large $\sigma$ without informative value.
    Instead we compute the relative frequencies 2048 times for 100 randomly chosen bricks ($b=16$), and obtain the standard deviation from these frequencies.}
    \label{fig:operation_frequency}
\end{figure}

\paragraph{Comparison to other Methods}
We compare the compression rate (CR) and time to the following methods (\cref{tab:comparison}):
\begin{itemize}
    \item hdf5: Segmented volumes are often provided in hdf5/Hierarchical Data Format~\cite{Motta:2019:DCR} which uses a brick-wise gzip-compression (LZ77 with Huffman coding); the brick size typically is $128^3$.
    \item png: Another common way is to slice volumes and store image stacks in Portable Network Graphics (PNG)-format where labels are split into 8-bit RGBA channels~\cite{Al-Thelaya:2021:TMG}. For our comparison, we slice the volumes along their z-axes and compress each 2D slice with the highest zlib compression level 9.
    \item Compresso~\cite{Matejek:2017:CEC}: This method has been designed for segmentation volumes and its CR often outperforms other techniques. We compare to the improved Compresso version 3.2. 
    \item Neuroglancer~\cite{neuroglancer}: A web-based volume viewer with a GPU-friendly compression format that stores a palette for each brick which is then indexed by the voxels.
\end{itemize}

\begin{figure}[tbp]
  \centering
    \includegraphics[width=\linewidth]{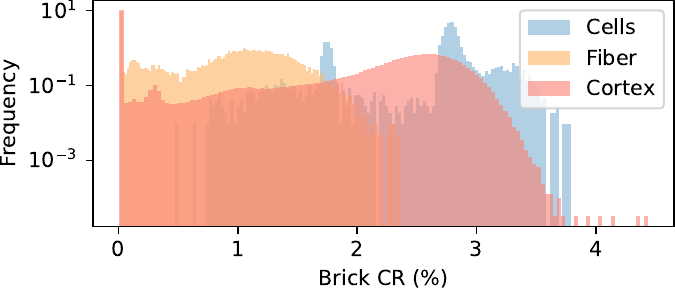}
    \caption{Log-scale histogram of compression rates per brick with $b=64$. We observe that even the worst case bricks still have compression rates of 3.794\% (\cells{}), 2.360\% (\fiber{}), and 4.436\% (\cortex{}). Note that the peak near zero for \cortex{} and \fiber{} is due to completely homogeneous bricks represented as a single $P_a$ entry.}
    \label{fig:cr_histogram}
\end{figure}

We can make the following observations: Compared to hdf5 our CSVs achieve better compression although we use smaller brick sizes.
Without LZMA, Compresso consistently yields worse CRs than our method without rANS.
Compresso with LZMA achieves roughly similar CRs as our method for \cells{} and \cortex{}. The result for the \fiber{}-volume is significantly worse than all other methods, which presumably is due to the strong anisotropy of the features and bad matching of representative windows used for compression in Compresso.
As an experiment we also used a global LZMA compression (as Compresso does) with the output of our brick encoding. By this we achieve slightly better CRs than Compresso (e.g. 2.579\% for ours and 2.753\% for Compresso on \cells{}). Note, however, that Compresso itself as well as using a single LZMA stream make level-of-detail rendering and brick-wise decompression impossible. This is why Compresso and image stack approaches are not directly applicable for volume rendering. hdf5-volumes, in contrast, can be decompressed per brick, but also have no elaborate level-of-detail mechanism.
Neuroglancer~\cite{neuroglancer} is meant for direct volume rendering and can also be trivially extended to store multiple levels-of-detail. However, already without LODs the compression rates are worse than with all other techniques.

As we can see, our CSV are competitive with the CRs of the state-of-the-art compression methods for segmentation volumes, and at the same time they can be directly used with volume rendering due to supporting brick-wise decompression and adaptive level-of-detail.

\cref{tab:comparison} shows \emph{single-threaded} compression time for all methods {as multi-threading implementations were not available for all methods}.
While the simple Neuroglancer-method encodes about twice as fast as our CSVs, its CR is significantly worse. Note that our CSVs are  $8-30\times$ faster than Compresso with LZMA while they achieve similar CRs (and significantly better CR for \fiber{}). Our rANS encoding introduces only little overhead.

\begin{figure}[tbp]
  \centering
    \includegraphics[width=\linewidth]{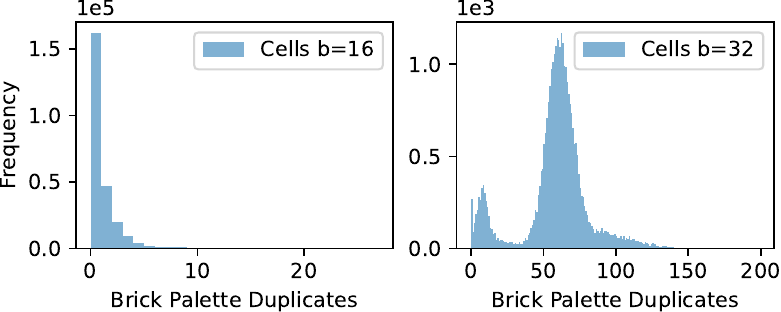}
    \caption{Histogram of the number of duplicate label entries in brick palettes for \cells{}. 
    On average a brick for $b=16$ has only 0.8 duplicate entries. Choosing $b=32$ results in 37449 bricks which contain 58 duplicates on average. This number remains relatively low also thanks to the $P_\delta$-operations. The \fiber{} and \cortex{} data sets have significantly fewer different labels per brick and duplicates are rare.}
    \label{fig:palette_duplicates}
\end{figure}

\paragraph*{Encoding Operation Frequencies}
\label{sec:res_frequency}
\cref{fig:operation_frequency} shows the relative frequencies of the encoding operations; note that they vary slightly, but not fundamentally, over the data sets.
We further observe that 95\% of the used operations reference other nodes ($R_p, R_{\{x,y,z\}}$), and only 0.08\% (\fiber{}) to 0.7\% (\cells{}) of nodes use the costly $P_a$ operation, adding an entry to the palette.
Recall that the frequencies of operations are also influenced by the order in which they are tested during encoding.  For example, $R_{\{x,y,z\}}$ would be more evenly distributed if tested in random order (which would harm compression).

As mentioned before we obtain the frequencies for encoding by sub-sampling bricks in a prepass. 
The cost for the prepass depends linearly on the number of sampled bricks and thus yields a significant speedup.
The compression rate with sub-sampled frequencies compared to a full pass over all bricks differs only on the fifth decimal digit, i.e. the frequencies vary only minimally over the bricks of the data set we tested.
In applications where a prepass is not practical, static (predetermined) frequency tables would still result in good compression ratios and they could still be tailored, for example, to a specific application (Cellular Potts Model, electron microscopy etc.).

In our tests, we observed data sets with bricks which showed noticeable worse compression ratio than the average -- however, the CR still remained below 5\%, which is less than the simple paletting as used by Neuroglancer (\cref{fig:cr_histogram}). We found that such bricks contain segmentation or simulation errors resulting in many disconnected label regions which inevitably lead to palette duplicates.
\cref{fig:palette_duplicates} shows histograms for duplicate palette entries for \cells{} (\fiber{} and \cortex{} exhibit almost no duplicates). Smaller brick sizes lead to fewer duplicates which is to be expected; larger bricks often span differently labelled regions and the required palette index of a node during encoding might not be adjacent and out of range for $P_\delta$ operations.

\begin{figure}[tbp]
    \centering
    \includegraphics[width=\linewidth]{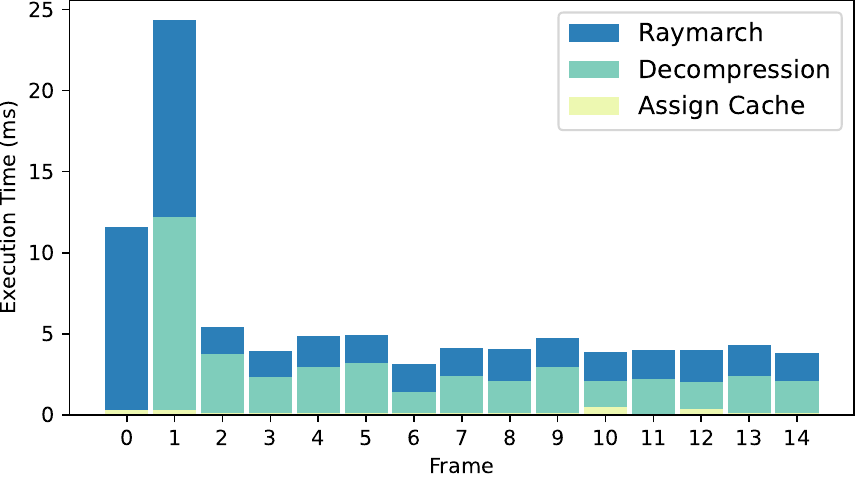}

    \vspace{0.1cm}
    \hspace*{0.7cm}\includegraphics[width=0.9\linewidth]{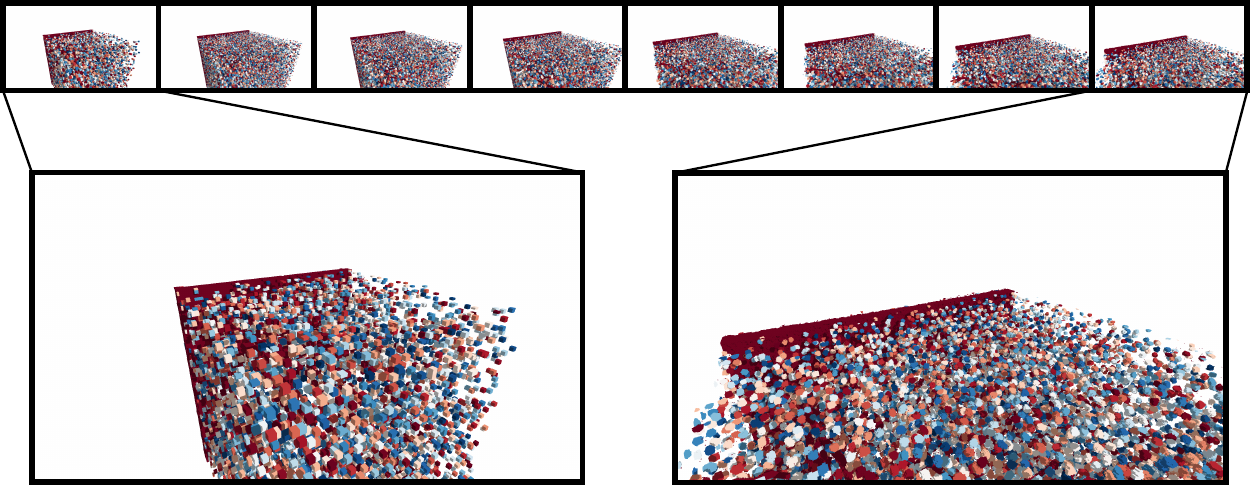}
    
    \caption{A stress test for the decompression with the \cells{} data set: a cold start, fast moving camera, and bricks becoming visible. We show the times for raymarching, decompression, and cache assignment at $1920\times 1080$ resolution. 
    The bottom row shows the individual frames (note how the coarsest LOD is used in frame 0 as no bricks are stored in the fresh cache). Changing brick LODs and changes in visibility also create decompression workload in subsequent frames.}
    \label{fig:times}
\end{figure}

\subsection{Rendering Performance}
\label{sec:renderingperformance}
We evaluate our raymarcher with the same hardware configuration as above using an NVIDIA RTX 3070 Ti with 8 GB of video memory. We use a cache size for decompressed bricks of 1 GB for \cells{} and \fiber{}; for these we can easily store the full CSV in video memory. For \cortex{} we set the cache size to 2 GB and separate the detail levels $L_0$ which consumes 9.2GB of CPU memory for all bricks. Recall that this detail is streamed to the GPU on-demand; the levels $L_l, l>0$ are kept in GPU memory and require 4.3 GB.
\Cref{tab:rendering} shows rendering performance with minimum, average, and maximum milliseconds (ms) per frame rendered at a resolution of $1920\times 1080$, and also lists the decompression throughput on the GPU.
\Cref{tab:rendering_bs} shows rendering performance using different brick sizes for \cells{} and \fiber{}.
Smaller brick sizes generally lead to slightly faster render times.
If not stated otherwise, we use $b=32$ as a compromise between rendering performance and compression, except for \cortex{} where we use $b=64$ for better compression and to minimize GPU memory usage.
As expected, the on-demand streaming of detail levels $L_0$ from CPU to GPU memory results in a slight reduction of rendering performance, but its influence is only notably evident for the maximum frame times. The maximum frame time for the large \cortex{} data set (172 ms) is due to a full cache rebuild (also see the supplemental video). The average render times as well as the maximum times for \cells{} and \fiber{} are only slightly affected by detail streaming. This is because in both cases all decompressed bricks easily fit into the cache.
For \fiber{} usually up to 600 MB of the cache are used at any point, while \cells{} requires less than 200 MB of the cache for decoded bricks.
With \cortex{}, we experience the cache quickly filling up to its 2 GB limit in close camera views when using transfer functions that lead to many visible bricks in the finest LOD. We leave more efficient caching schemes for future work.

\cref{fig:times} shows the total frame times as well as the portions spent for raymarching, decoding, and cache assignment for a synthetic test using \cells{} designed to put stress on the decompression by fast camera movement and change in the visibility of bricks. The short sequence consists of 15 frames where the cache is empty in the beginning. Consequently frame 0 is rendered using the coarsest LOD only and significant decompression load is generated for frame 1. Even the initial decompression of all bricks visible in frame 0 results in a total frame time below 25 ms. Note that 
 similar cases in practice only occur once at the beginning of rendering a sequence or when the camera view changes (almost) completely. The subsequent frames in the experiment still require further decompression due to bricks becoming visible and selection of LODs, but yield roughly equal total frame times below 5ms.
Also note that the cache assignment stage, which among others checks a brick's visibility by applying the transfer function to the palette, only accounts for a negligible overhead.

\begin{table}[t]
\centering
\caption{Average, minimum, and maximum total rendering times per frame in milliseconds (ms) for a fly-around at $1920\times 1080$ resolution, and GPU-decompression performance in GB/s for our data sets. The render times include one shadow ray per pixel. First column: rendering with CSVs completely stored in GPU-memory; second column: CSV split such that $L_0$ is stored in CPU-memory.}
\label{tab:rendering}
\begin{tabular}{r|l|l|l}
& CSV on GPU & $L_0$ on CPU & decoding\\
& min / avg / max & min / avg / max & in GB/s\\
\midrule
\cells{} &  5 / 17 / 29  &  5 / 20 / 39 & 9.9\\
\fiber{} & 7 / 18 / 23 & 7 / 22 / 30 & 10.4\\
\cortex{} & {\centering --} & 7 / 40 / 172 & 9.3\\
\end{tabular}
\end{table}%
\newcommand{\timing}[3]{#2 / #3}
\begin{table}[t]
\vspace{0.1cm} 
\centering
\caption{Average and maximum frame times in milliseconds with varying brick sizes $b$ for a fly through over \cells{} and \fiber{}. 
\cortex{} cannot be rendered with $b < 64$ on our GPU with 8 GB memory.}
\label{tab:rendering_bs}
\begin{tabular}{r|ccc|ccc}
& \multicolumn{3}{|c}{CSV on GPU} & \multicolumn{3}{|c}{$L_0$ on CPU} \\
$b$ & 16 & 32 & 64 & 16 & 32 & 64\\
\midrule
\cells{} & \timing{3}{6}{12} & \timing{5}{17}{29} & \timing{5}{23}{37} & \timing{3}{9}{15} & \timing{5}{20}{39} & \timing{7}{24}{43}\\
\fiber{} & \timing{5}{10}{13} & \timing{7}{18}{23} & \timing{14}{33}{39} & \timing{5}{12}{15} & \timing{7}{22}{30} & \timing{14}{35}{39}\\
\end{tabular}
\end{table}%

\section{Conclusions}
We presented a novel lossless compression technique for voxel-based segmentation volumes which achieves compression ratios comparable to, or better than the state-of-the-art. At the same time, its brick-wise compression provides sufficient granularity for efficient volume visualization, and the multi-resolution encoding inherently enables decompression with adaptive level-of-detail. We have demonstrated volume visualization using raymarching and caching of decompressed bricks for data sets with more than 900 GB on modest hardware at real-time frame rates. 
We have also outlined possible venues for future work to further improve the performance by increasing the utilization of CPU cores, improved empty-space skipping, predictive decompression, or caching. Even without these optimizations our method achieves high throughput for the compression and real-time visualization of very large segmentation volumes.


\section*{Supplemental Materials}
\label{sec:supplemental_materials}
Please see (1) the accompanying video showcasing our technique and (2) the source code of our compression method released under a \mbox{CC BY-NC 4.0} license. We also provide an (3) in-depth comparison of using Morton over Hilbert curves in the compression.




\acknowledgments{%
The authors wish to thank the NIC Research Group Computational Structural Biology at Jülich Research Center for simulating and providing the Cellular Potts Model data sets and for their helpful feedback.
We also wish to thank the Research Group Computed Tomography, University of Applied Sciences Upper Austria, Campus Wels, where the Fiber data was measured and analyzed for providing this data set.
This work has been supported by the Helmholtz Association (HGF) under the joint research school ``HIDSS4Health – Helmholtz Information and Data Science School for Health'' and through the Pilot Program Core Informatics.
}

\bibliographystyle{abbrv-doi-hyperref}

\bibliography{neighborhood_compression}









\end{document}